# Precision Thermo-Welding of Polymer Microspheres into Periodically Organized, Hybrid, Mechanically Rollable Coupled-Resonator Optical Waveguides

*Kapil Yadav,[#] Jyotisman Hazarika,[#] Mehdi Rohullah, Melchi Chosenyah, Rajadurai Chandrasekar\**

K. Yadav,[#] J. Hazarika,[#] M. Rohullah, M. Chosenyah, Prof. R. Chandrasekar

School of Chemistry and Centre for Nanotechnology, University of Hyderabad,

Prof. C. R. Rao Road, Gachibowli, Hyderabad 500046, India

E-mail: r.chandrasekar@uohyd.ac.in

[#] Equal contribution of authors

## Abstract

Polymer microspherical resonators that trap light are crucial structures for on-chip and on-board integration in nanophotonic applications. Using advanced micromanipulation and thermo-welding methods, we successfully create welded polystyrene-based coupled-resonator optical waveguides (CROWs) with customized lengths, shapes, and optical characteristics. Through the thermal fusion of blue, green, and red fluorophore-doped polystyrene microspheres, multi-fluorescent cohesive units are created from dimeric to henicosameric periodic arrangements. Detailed micro-spectroscopy studies reveal CROWs' capability to support optical whispering-gallery modes. The periodic arrangements of different fluorophore-doped resonators within the CROW facilitate efficient guided transmission of light *via* both active and passive mechanisms in opposite directions. The welded structures exhibit mechanically driven rolling motion, confirming their integration as cohesive units. These findings highlight the versatility and performance of thermo-welded polymer microsphere-based waveguides, paving the way for scalable, low-cost photonic devices in sensing, modulation, and light guiding, emphasizing their role in future integrated photonics.

**1. Introduction**

Self-assembled polymer optical materials are receiving increasing attention for their potential in optoelectronics and nanophotonics, driven by their mechanical flexibility, tunable optical properties, lightweight nature, relatively high refractive index (around 1.6), non-linear optical (NLO) property, and ease of solution and melt processing.[1-4] Among these materials, polystyrene (PS) stands out due to its ability to form spherical microparticles of varying sizes through solvent-assisted self-assembly. In the field of nanophotonics, the spherical geometry of microparticles facilitates efficient internal circulation of trapped light due to their high refractive index and smooth surfaces.[3,5,6] This makes them particularly well-suited for the development of high-quality ($Q$) factor whispering-gallery-mode optical resonators (WGMRs), which are essential for lasing, high-precision sensing, imaging, and signal processing applications.[6] The self-assembly of uniform-sized PS particles also plays a key role in the formation of photonic crystals.[7] However, until now, only discrete optical microparticles have been utilized for various light-based applications.[5,6] PS-based optically linear and NLO-WGMRs with different emission band width of the electromagnetic spectrum can be prepared by doping the PS with different dyes.[5a-c] Highly π-conjugated co-polymer and polydimethylsiloxane based WGMRs demonstrate lasing at a low pump-threshold.[6] Yamamoto *et al.* have demonstrated photochemically switchable PS WGMRs interconnected with polymer microfibre for all-organic optical logic gate applications.[8]

Recently, we demonstrated the precise mechanical micromanipulation of two PS WGMRs into a dimer (not welded) using an atomic force microscopy (AFM) cantilever on a coverslip by following mechanophotonic approach.[9,10b] Our group reported the possibility of performing photon molecular reactions by assembling dimeric and L-shaped polymer WGMRs.[10b] In the next stage of advancement, these microcavities can be manipulated into precisely arranged geometries to physically weld them. By linking them together, a permanently fused, chain-like coupled-resonator optical waveguide (CROW) can be created,[11] which efficiently guide light through evanescent coupling, even around sharp angles. Heating PS microspheres that are in physical contact above their glass transition temperature ($T_g \approx 100$ °C)[12] causes them to weld at the contact points. There is a lack of reports on the fabrication of well-defined one-dimensional (1D) architectures formed by micro-welding of PS microparticles with varying lengths and geometries. The creation of such innovative photonic architectures depends on the precise control over the arrangement of PS microparticles, followed by their welding. However, the physical assembly and subsequent welding of dye-doped microresonators into permanently connected structures of varying lengths and periodic arrangements for waveguiding applications, has not been demonstrated. This fused architecture is crucial for the development of both monolithic and hybrid (multiple FL compound-doped) PS CROWs, which have significant potential for advanced applications in optoelectronics and photonics.[13] Furthermore, the precise manipulation of these CROWs and their mechanical rolling locomotion presents a fundamental challenge in realizing smart polymer-based photonic devices.

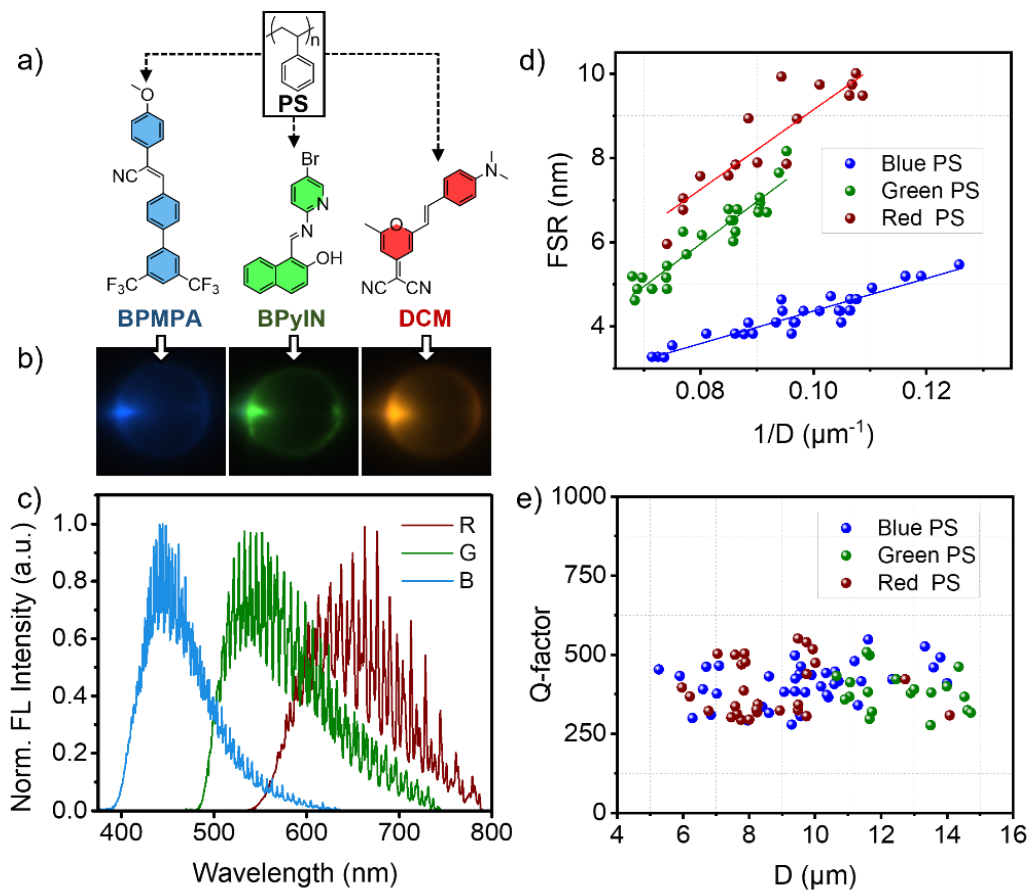

**Figure 1.** a) Schematic illustration of the formation of blue, green, and red emitting PS optical microcavities from the self-assembly of PS with BPMPA, BPyIN, and DCM molecules, respectively. b) FL microscopy images of the doped PS WGMRs and their c) FL spectra. d) A plot of FSR versus 1/D for the doped microcavities. The lines show the linear fits. e) A plot of $Q$-factor versus particle D for the microcavities.

This innovative work showcases the realization of monolithic and hybrid thermo-welded dimeric, trimeric, tetrameric, decameric, undecameric and henicosameric (in Greek *henicos-* means 21) CROW fluorophore-doped PS CROWs with pre-determined lengths, color combinations, and geometries, all designed to enable mechanically powered rolling locomotion. These CROWs are created using a mechanophotonics approach by precisely assembling PS microspheres doped with blue, green, and red-emitting fluorophores in different sequences, allowing the fabrication of monolithic and hybrid fluorescent CROWs with tunable emission across the visible spectrum. Integrating diverse fluorophores into a single hybrid CROW structure enables efficient passive and active light propagation via optical energy transfer and/or evanescent coupling, leveraging the distinct absorption and emission properties of each fluorophore. These welded PS CROWs exhibit excellent mechanically triggered rollability. Additionally, multiply bent fused CROW architectures of varying fluorophore-doped polymer microcavities can be made, making them promising candidates for applications in innovative soft photonic devices.

**2. Results and discussion**

Initially, blue, green, and red-fluorescent organic molecules namely, (*Z*)-3-(3',5'-bis(trifluoromethyl)-[1,1'-biphenyl]-4-yl)-2-(4-methoxyphenyl)acrylonitrile (BPMPA), (*E*)-1-(((5-bromopyridin-2-yl)imino)methyl)naphthalen-2-ol (BPyIN), and (*E*)-2-(2-(4-(dimethylamino)styryl)-6-methyl-4H-pyran-4-yl)malononitrile (DCM), respectively were prepared employing reported literature (Scheme S1, S2, Supporting Information).[14] The fluorophore-doped polystyrene microspheres were obtained from the molecular self-assembly technique of their respective organic molecules. (**Figure 1**a; see the experimental section). Confocal optical microscope images confirmed the successful incorporation of the compounds within the microspheres, as evidenced by the corresponding fluorescence (FL) observed upon excitation with a 405 nm continuous-wave (CW) laser (Figure 1b). Upon laser excitation at the edge of the doped microspheres, the corresponding FL spectra appeared in the following regions: approximately 390-610 nm (blue), 480-740 nm (green), and 540-800 nm (red). Further, the spectra exhibited multiple pairs of sharp peaks, indicating the formation of WGMs due to the spherical geometry of the cavity (Figure 1c). Additionally, the observed peak pairs were found to correspond to transverse electric and transverse magnetic modes. The free spectral range (FSR) (representing the separation between adjacent peaks of the same mode number, $m$) or $\Delta\lambda = \lambda_m - \lambda_{m+1}$, were analyzed for multiple microspheres of varying diameters (D). The FSR value is inversely related to the D of the microcavities as per equation, FSR (or) $\Delta\lambda = \frac{\lambda^2}{\pi D n_{\text{eff}}}$, where $n_{\text{eff}}$ is the effective refractive index. The nearly linear relationship observed in the FSR versus 1/D plot (Figure 1d) confirms the diameter-dependent behavior of these doped polystyrene WGMRs. The *Q*-factor is a parameter that symbolizes the light trapping effectiveness of a microresonator which is described as $Q = \lambda/\Delta\lambda$, where $\Delta\lambda$ is the line width of a peak at full width at half maximum (FWHM) and $\lambda$ is the wavelength of the peak. The plot for *Q*-factor versus D showed that the *Q*-factor is in the range of ≈250-550 for various microcavities (Figure 1e). The variation in the *Q*-factor for microcavities of the same diameter indicated the presence of differing levels of optical losses among them.

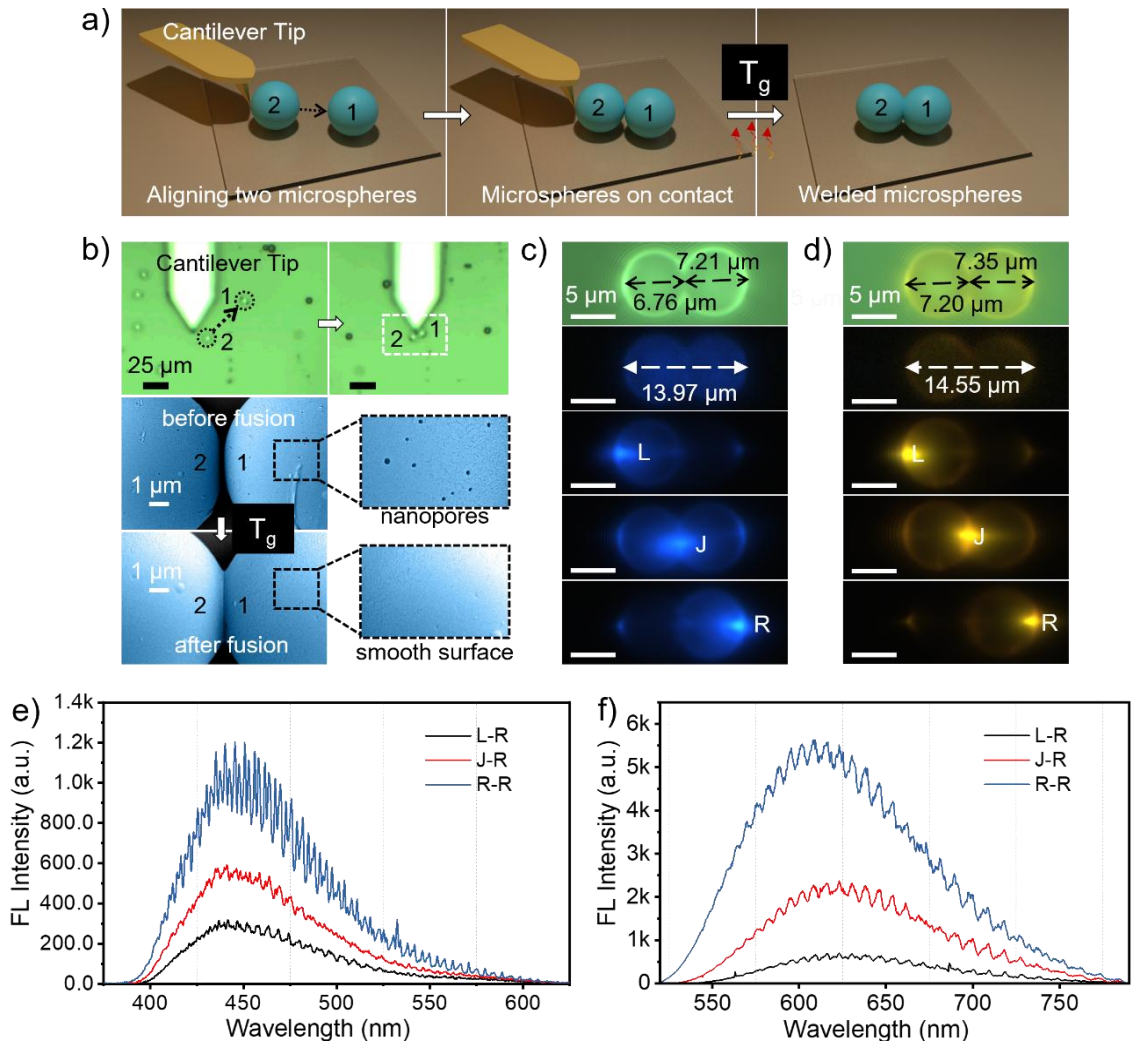

**Figure 2.** a) Schematic illustration of mechanically assisted construction of a dimeric CROW using an AFM cantilever tip. b) Sequential confocal optical images of constructing dimeric CROW and its corresponding false color-coded FESEM images before and after fusion. The framed images show the difference in surface morphology of the PS microspheres before and after fusion. Confocal optical and FL microscopy images of c) BPMPA-doped and d) DCM-doped dimers for various excitation points with a 405 nm laser and their respective e, f) FL spectra. L-R, J-R, and R-R stand for excitation at the left, joint, and right, and signal collection at the right side of the dimer.

To create stable dimeric microspheres doped with blue-emissive BPMPA, one microsphere of 6.76 μm diameter was mechanically pushed towards another of D = 7.21 μm using an AFM cantilever tip to establish physical contact. The evanescent coupling in the assembled structure was studied by optically exciting the edges of the microspheres and recording their FL spectra at different positions labeled left (L), joint (J), and right (R) and recording the guided signals at R (Figure S1, Supporting Information). Subsequently, the substrate containing the microspheres was heated to its $T_g$ in a tailored heating environment (Figure S2 for the experimental set-up, Supporting Information). This facilitated the thermo-welding of the microspheres at the contact region due to the inter-diffusion of polymer chains (**Figure 2**a). The temperature and heating interval were optimized to soften the microspheres and form a neck-like bridge in the contact region while preventing localized heat-induced deformation. When the temperature is reduced, the fusion between the microspheres is retained due to the thermoplastic nature of PS. The dimerization of the microspheres was confirmed through the unified mechanical movement of the structure when pushed through an AFM cantilever tip attached to a confocal optical microscope and further affirmed through FESEM (Figure 2b). The electron microscopy images post-fusion also revealed the smoothened morphology of the microcavities and a narrow reduction in the length of the CROW. Optical microspectroscopy investigation of the welded microspheres revealed that the dimeric microspherical resonators are also optically coupled, resulting in the formation of WGMs in the FL spectra (Figure 2e). Similar welding was also achieved in red-emissive DCM-doped microspheres (D=7.20 μm and 7.35 μm), resulting in WGM resonances. Notably, excitation at one end of these dimeric microspheres guided light to the other end of the dimer (Figure 2d).

The corresponding FL spectra exhibited the signature of WGM, confirming that the dimer functions as a CROW (Figure 2f).

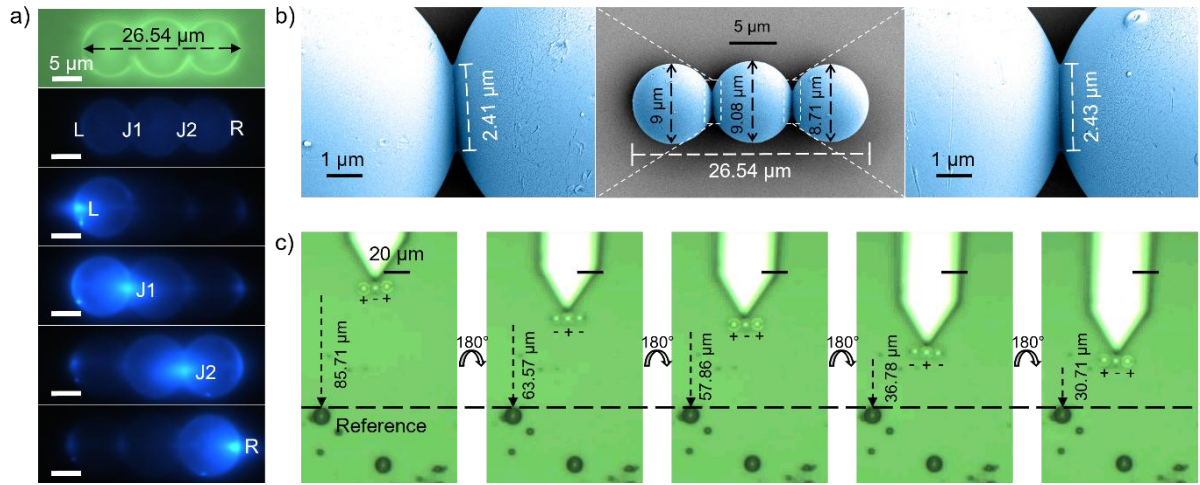

**Figure 3.** a) Confocal optical and FL microscopy images (405 nm laser excitation) of trimeric CROW. b) Color-coded FESEM images of the trimeric CROW. c) Sequential confocal optical images of rolling the CROW with an AFM cantilever tip. '+' and '-' represent the bright and dark contrast in the z-axis.

The rolling locomotion of one-dimensional organic crystals is typically hindered due to their non-circular cross-sections. Contrariwise, spherical microcavities, due to their geometry, can be exploited for rolling using precise micromanipulation techniques. This strategy can be implemented to design monolithic and hybrid rectilinear CROWs capable of functioning as efficient one-dimensional rotors with forward and backward mobility. To demonstrate this, about ≈26.54 µm long BPMPA doped fused trimeric CROW was fabricated with precise micromanipulation techniques by identifying three microspheres of nearly comparable sizes (8.71 µm, 9.08 µm, and 9 µm) and placing them adjacent to each other (Figure S3, Supporting Information). The arrangement was thermo-welded to yield the desired fused trimeric CROW (**Figure 3**a), and the amalgamation was confirmed through FESEM (Figure 3b). The close-up FESEM images of the trimeric CROW revealed that the width of the welding regions is about 2.4 µm. For a standard rolling experiment, the fabricated CROW was placed under a confocal optical microscope equipped with an AFM cantilever tip. The trimeric CROW was initially subjected to mechanically-assisted translational motion on the substrate to confirm cohesion by pushing it with a cantilever tip. Subsequently, a gentle forward mechanical force was applied to the center of the fused trimeric CROW with the cantilever tip, enabling the structure to roll smoothly from its initial position of 85.71 µm away from the reference point to a position of 63.57 µm from the reference. This motion led to a reversal in contrast (due to a slight variation in microsphere height), with the bright region becoming dark and *vice versa*, thus signifying a 180º roll on the substrate. The structure was then stepwise moved to various target positions, attaining distances of approximately 57.86 µm, 36.78 µm, and 30.71 µm distant from the reference point (Figure 3c, Supporting Video 1). For each translation, the contrast of the microcavities alternated between bright and dark. These observations provide clear evidence of the rolling motion of the rectilinear fused CROWs.

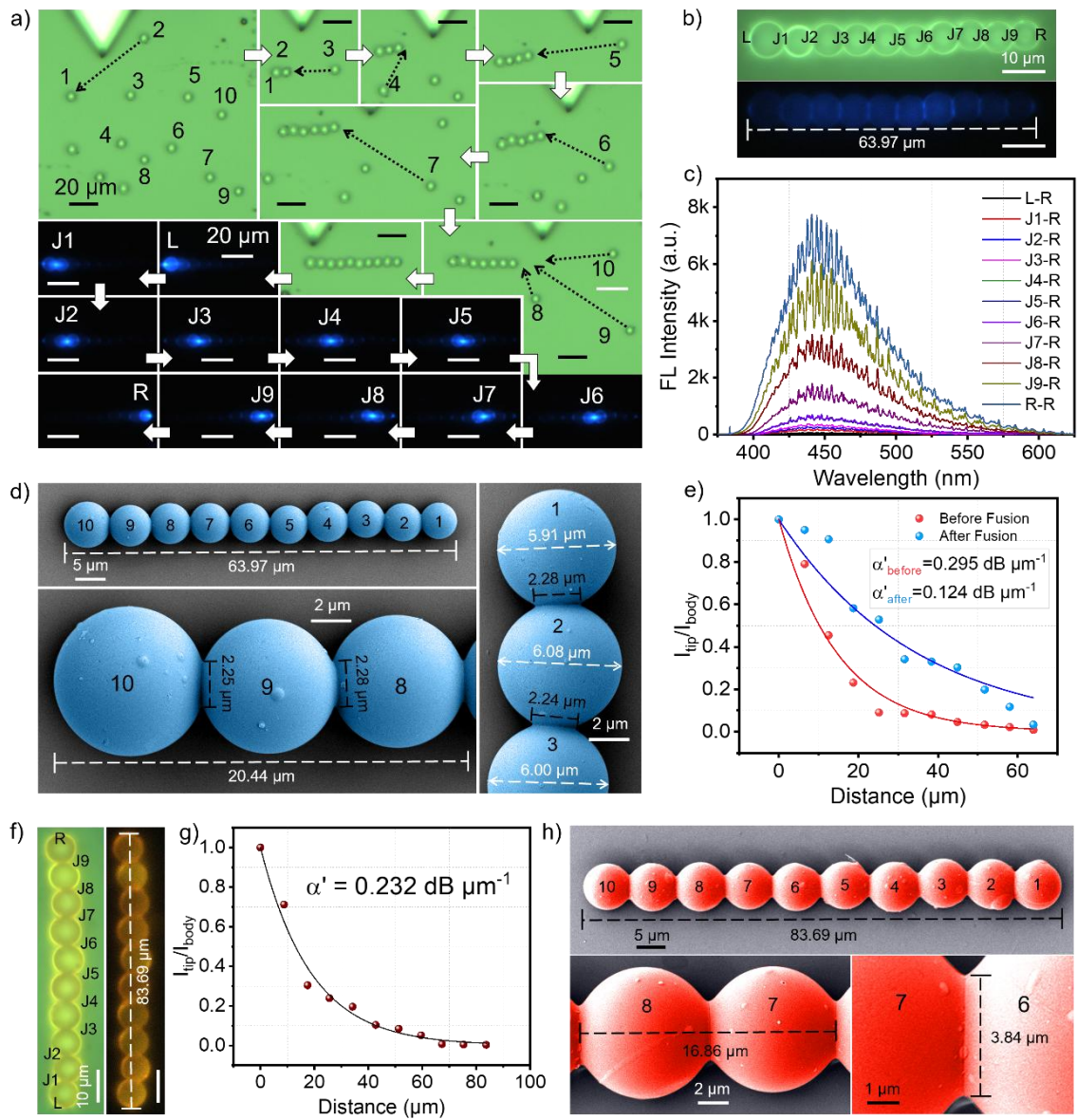

**Figure 4.** a) Sequential confocal optical images of constructing a BPMPA-doped CROW and its FL microscopy images for different excitation points with a 405 nm laser. b) Magnified confocal optical and FL microscopy images of the BPMPA-doped CROW and its corresponding c) FL spectra recorded before fusion. d) False color-coded FESEM images of the BPMPA-doped CROW after fusion. e) The plot of $I_{tip}/I_{body}$ versus distance for the BPMPA-doped CROW used for the comparison of the optical loss before and after fusion. f) Confocal optical and FL microscopy images of a DCM-doped CROW after fusion and the corresponding g) plot of $I_{tip}/I_{body}$ versus distance used for the calculation of the optical loss. h) False color-coded FESEM images of the DCM-doped CROW after fusion.

Using the same approach, one-dimensional rectilinear monolithic CROWs of varying lengths, ranging from 26.5 µm (trimer) to 44.4 µm (tetramer), were constructed with different numbers of PS microspheres doped with BPMPA and DCM molecules. Their optical waveguiding properties were then studied (Figure S4, Supporting Information). The optical excitation at the extreme left edge within the CROW structures generated intense FL at the extreme right edge, along with their characteristic WGM spectra.

We extended this strategy to construct a CROW consisting of ten blue BPMPA-doped PS microcavities to show successful long-distance light-guiding capability and mechanically assisted locomotion. A region comprising microspheres with comparable sizes was identified, and a decameric CROW was assembled in a stepwise manner with precise movements using an AFM cantilever tip (**Figure 4**a). The assembled structure was probed with confocal optical microscopy to ensure each microsphere was in contact with its adjacent microsphere. Thermo-welding of the geometry was carried out to form a unified decameric CROW. The unification of the CROW structure was confirmed through FESEM (Figure 4d). Optical excitation at the extreme left edge of the CROW geometry exhibited propagation of blue FL to the right edge of the tenth microsphere. The FL spectra displayed the signature WGMs. Similarly, the laser excitation was translocated to the adjacent junctions in a sequential manner, and the FL recorded at the extreme right edge sustained its WGMs at all instances (Figure 4c). Further, the optical loss coefficient ($\alpha$) before and after welding was estimated to be 0.295 dB μm$^{-1}$ and 0.124 dB μm$^{-1}$ using the following equation.

$$\frac{I_{tip}}{I_{body}} = e^{-\alpha' L}$$

where $I_{tip}$ and $I_{body}$ correspond to the FL intensity collected at the tip and the excitation point of the waveguide, respectively, and $L$ is the optical path length. From the fit value of $\alpha'$, optical loss $\alpha$ in dB μm$^{-1}$ (dB loss = 4.34 $\alpha'$) was estimated (Figure 4e). Evidently, the current strategy outperforms the traditional CROWs at efficient optical transduction. Additionally, the fused CROW was subjected to AFM cantilever tip-assisted mechanical movements along its long axis, and it successfully retained its geometry in all respects (Figure S5, Supporting Information). Similar experiments were carried out for a decameric CROW with red-emissive DCM-doped microcavities, and they yielded analogous results (Figure S6, Supporting Information; Figure 4f-h).

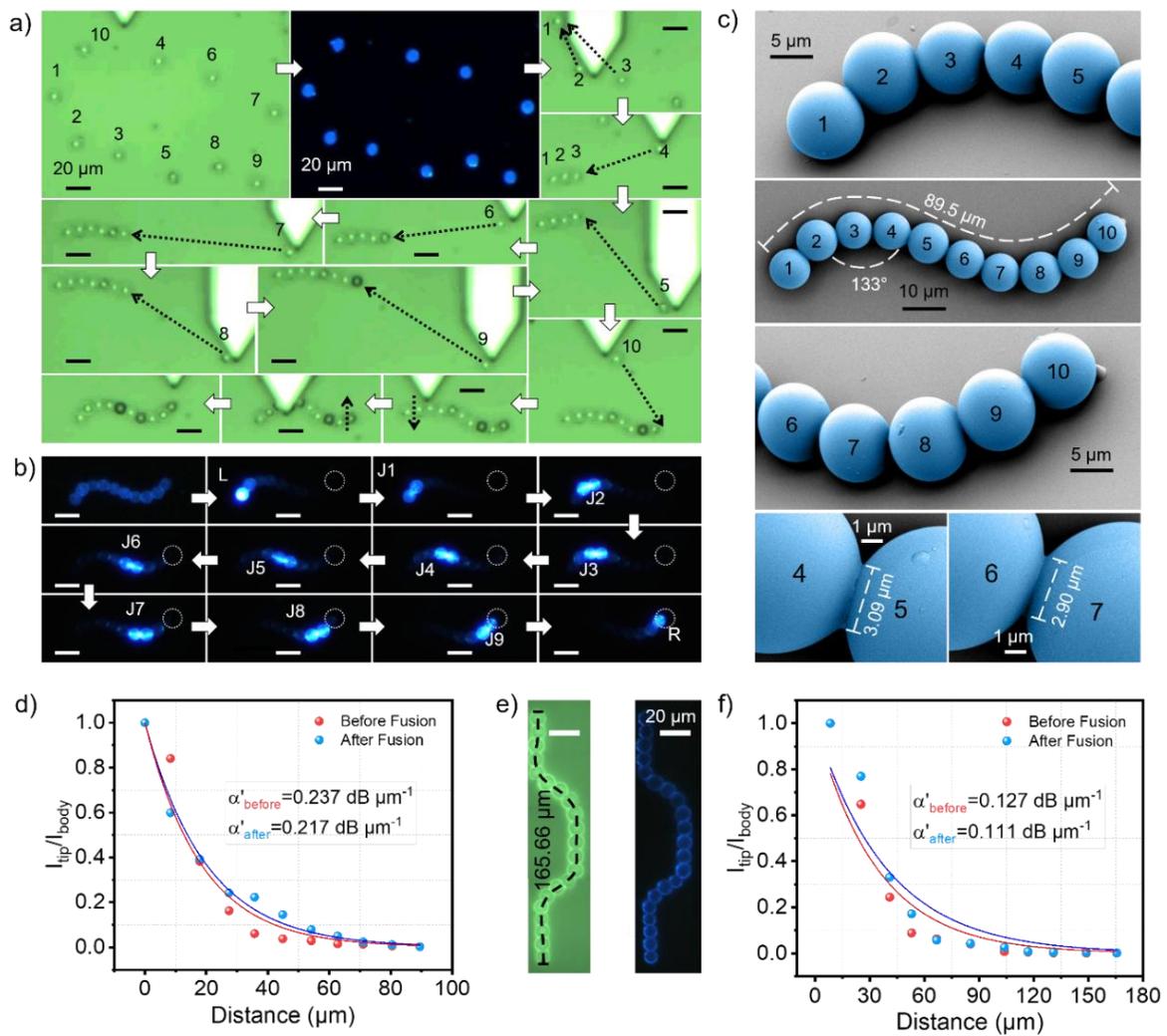

**Figure 5.** a) Sequential confocal optical images of constructing a BPMPA-doped doubly bent decameric CROW and its b) FL microscopy images for different excitation points with a 405 nm laser. c) False color-coded FESEM images of the doubly bent geometry after fusion and its d) plot of $I_{tip}/I_{body}$ versus distance used for the comparison of the optical loss. e) Confocal optical and FL microscopy images of a BPMPA-doped henicosameric CROW with four bends and the corresponding f) plot of $I_{tip}/I_{body}$ versus distance.

As curved optical transduction paths are useful for circuit technologies to route the light in clockwise or counterclockwise directions, we fabricated a decameric CROW with a doubly bent geometry. Ten microspheres of comparable sizes were identified and mechanically brought into contact with an AFM cantilever tip in a sequential approach to form a ≈89.5 µm long CROW with a curved geometry with a curvature of ≈133º (**Figure 5**a). After the photonic experiments to assess the *α* value (Figure 5b), the curved CROW was subjected to thermo-welding, which resulted in a CROW with a permanently fused doubly bent geometry. The fusion was confirmed through FESEM, and it revealed a consistent welding width of nearly 3 µm at all the junctions (Figure 5c). The optical studies disclosed a reduction in optical loss of 0.020 dB µm$^{-1}$ after the welding of the doubly bent geometry, thus enhancing the light-guiding efficiency in the curved CROW structure. (Figure 5d). Additionally, a ≈165.7 µm long thermo-welded henicosameric CROW with four bends was achieved (Figure 5e). The optical studies revealed an analogous optical loss decrement of 0.016 dB µm$^{-1}$ post-welding, thus establishing the efficiency of the current strategy (Figure 5f).

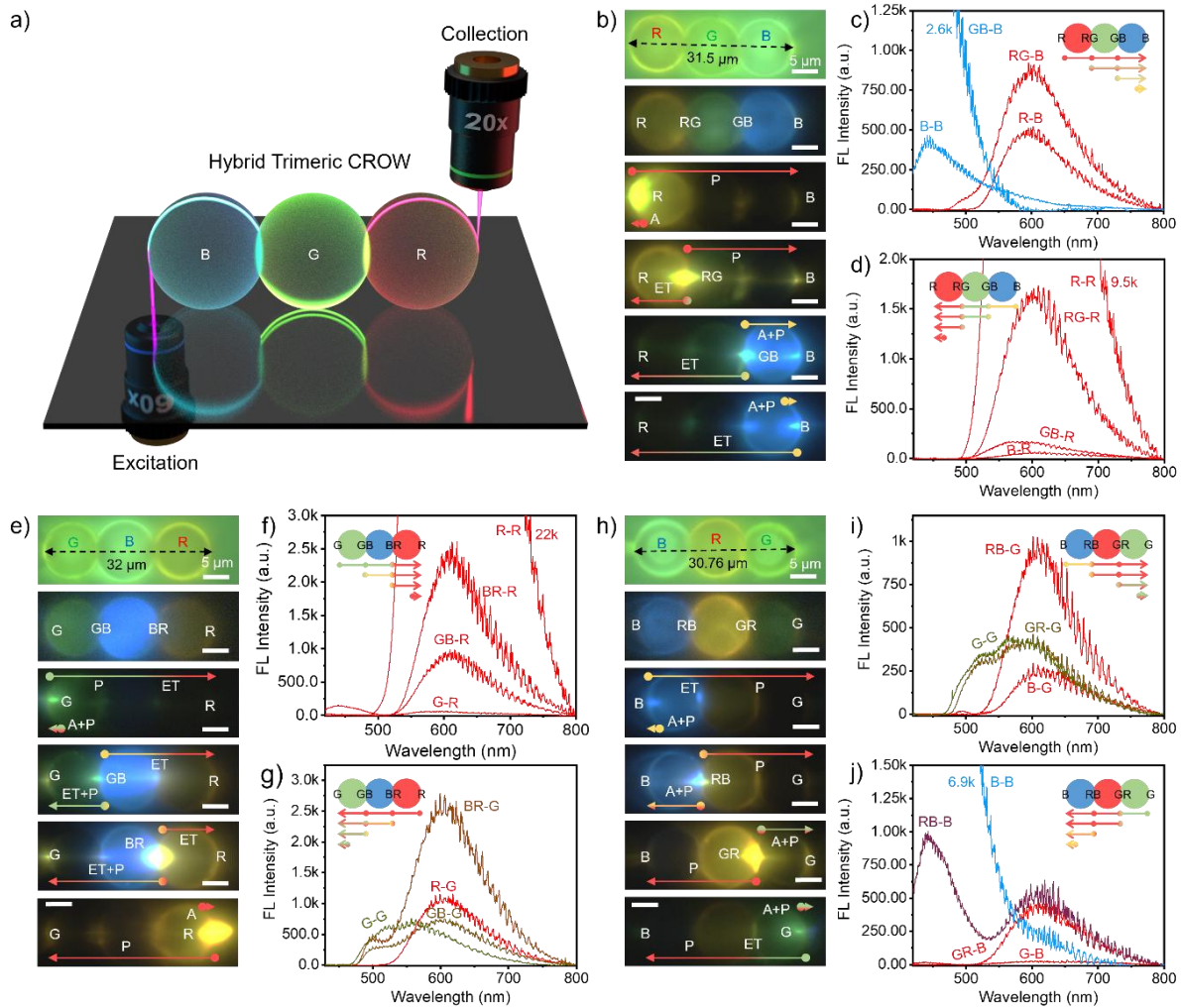

**Figure 6.** a) Graphical illustration depicting optical propagation of a hybrid trimeric CROW. b, e, h) Confocal optical and FL microscopy images of hybrid trimeric CROWs and their corresponding c, d, f, g, i, j) FL spectra. A, P, and ET stands for active waveguiding, passive waveguiding, and energy transfer, respectively. In the FL images, the brightness and contrast were adjusted by +40% and -40%, respectively. The original images are shown in Figure S7, Supporting Information, for clarity.

We expanded the mechanical micromanipulation technique to assemble microcavities doped with different organic fluorophores, arranged in varying color sequences, into one-dimensional rectilinear trimeric hybrid CROWs (**Figure 6a**). We identified PS WGM microcavities doped with red (R), green (G), and blue (B) fluorophores of similar sizes, and arranged them in serial combinations of R-G-B, G-B-R, and B-R-G CROWs of length 31.5, 32 and 30.8 µm, respectively (Figure 6b, e, h). For the R-G-B trimeric CROW (Figure 6b–d), excitation at R with a 405 nm CW laser generated red light at B *via* a passive mechanism, as no energy transfer occurs from R to G or from G to B in the R-G-B arrangement (Figure 6b). When optical excitation was applied at the RG junction, a prominent red FL with a green FL shoulder was detected at B through a passive light-guiding mechanism (Figure 6b,c), while red light was observed at R *via* an active mechanism (Figure 6b,d). Excitation at the junction GB resulted in passive outcoupling of green and blue FL bands towards B (Figure 6b,c), with an active red signal output at R (Figure 6b,d). The active mechanism observed is due to sequential light propagation from GB to R through G, driven by evanescent coupling-assisted energy transfer (ET). This process is driven by the overlapping emission and absorption spectra between adjacent microcavities in the B-G-R CROW (Figure 6c; Figure S8, Supporting Information). Similarly, reverse excitation at edge B generated a bright red FL at edge R within the 520–780 nm spectral range, demonstrating an active waveguiding mechanism facilitated by ET from B to G and then to R microspheres (Figure 6b,d). This experiment demonstrates the ability of the constructed CROW to guide the same red light to the output, irrespective of whether excitation occurs on the extreme right or left side of CROW, utilizing two distinct light-guiding mechanisms.

The optical excitation and signal collection at the edge of G in the G-B-R trimeric CROW showed a major green emission (Figure 6e,g). For the same excitation, the recorded signal at the edge of R exhibited a red signal due to the passive mechanism (Figure 6e,f). This confirms that the green signal propagates passively through resonator B, thereby exciting resonator R *via* ET. Exciting junction GB resulted in a strong red signal and a weak blue signal as output in resonator R, suggesting significant energy transfer of the blue emission to excite the red-emissive resonator (Figure 6e,f), while a dual green and red band was recorded at the edge of G (Figure 6e,g). Similarly, excitation at junction BR mainly produced a red signal at resonator R (Figure 6e,f) and a strong red signal along with a weak green signal at edge G (Figure 6e,g). When the edge of resonator R was excited and collected at the same edge, a strong red signal was recorded (Figure 6e,f). For the same excitation, a strong red signal was recorded at the edge of the G resonator, confirming the passive propagation of the red signal through the R to the G resonator (Figure 6e,g).

Excitation and spectral collection at the edge of B of the B-R-G trimeric CROW showed a strong blue signal and a weak red signal due to the excitation of the neighboring R resonator by the B resonator *via* ET and subsequent passive propagation of the red signal through B resonator (Figure 6h,j). Optical excitation of the junction RB generated a mixed spectrum covering the blue and red bands at B and passive propagation of the red signal to G (Figure 6h–j). Likewise, excitation at junction GR delivered a strong red signal at B *via* passive mechanism and a mixed green and red signal at G *via* active and passive mechanisms (Figure 6h–j).

Excitation of the edge of G produced a guided weak green and red signal at B. The FL spectra were recorded for all the possible excitation-collection points on the hybrid trimers' body, which validated the optical waveguiding nature of the fused multi-fluorescent CROWs obtained using the proposed strategy.

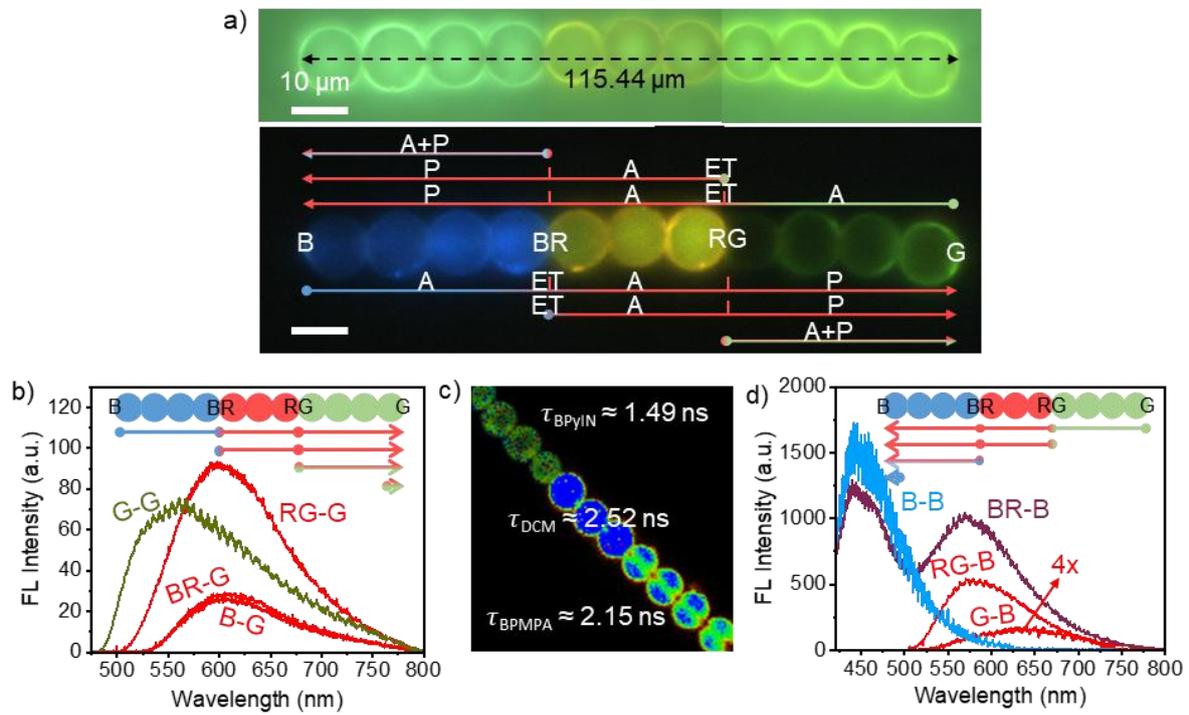

**Figure 7.** a) Stitched confocal optical and FL microscopy images of hybrid undecameric (4B+3R+4G) CROW and its corresponding b, d) FL spectra, and c) FL-lifetime image. ET, P, and A stands for energy transfer, passive waveguiding, and active waveguiding, respectively.

Further, a hybrid, linear, welded 4B+3R+4G, undecameric CROW of length 115.4 µm was made by using eleven PS WGMRs (Figure 7a). This undecameric CROW also displayed efficient photonic transport from red to blue regions depending upon the excitation position with minimal optical losses (Figure 7b). For example, upon excitation at the edge of the blue trimer (B), the emitted blue light actively propagates toward the red trimer, exciting it via evanescent coupling assisted ET. The resulting red light is then again actively propagated towards the RG junction. However, within the green trimer (G), where no ET occurs, the red light passively propagates to the edge of G (Figure 7b). Similarly, excitation of either BR or RG junctions delivered a red signal guided to G via active/passive and passive mechanisms, respectively. Upon optical excitation at the BR junction, a strong red FL was observed at the edge of G, driven by ET at the BR junction, followed by active waveguiding through the red microcavities and passive waveguiding through the green microcavities. Under the same excitation, a mixed FL spectrum of blue and red was detected at the edge of B, as confirmed by the double-humped FL spectrum with $\lambda_{max}$ at 440 nm and 570 nm (Figure 7d). In this process, the red signal propagates via passive waveguiding, while the blue signal follows an active waveguiding mechanism.

Likewise, excitation at the RG junction generated a mixed broad spectrum (500-800 nm) covering the red and green bands at the edge of G, with passive propagation of the red signal and active propagation of the green signal. The same excitation also induced a red FL signal at the edge of B *via* ET transfer at the junction, followed by the active waveguiding mechanism through the red microcavities and the passive waveguiding mechanism through the blue microcavities. Excitation at edge G produced a weak red signal at edge of B, generated *via* an active waveguiding mechanism through the blue cavities, followed by ET at the junction, with an active propagation of the signal through the red microcavities and passive waveguiding mechanism through the blue microcavities (Figure 7b,d).

To investigate the FL lifetime properties of the multi-fluorescent hybrid undecameric CROW geometry, FL-lifetime imaging microscopy (FLIM) studies were performed. The FLIM images of a selected area (80×80 µm$^2$) showed an average FL decay lifetime value of 1.49 ns, 2.52 ns, and 2.15 ns for BPyIN, DCM, and BPMPA-doped PS WGMRs, respectively (Figure 7c; Figure S9, Supporting Information). This observed lifetime aligns well with the expected emission characteristics of the incorporated fluorophores, suggesting successful doping and uniform dispersion within the polymer microcavities.

## 3. Conclusion

This study presented the remarkable potential of innovatively fusing polymer microcavities into CROWs, advancing the field of photonics with their inherent structural versatility and high-efficiency multicolor light transmission capabilities. The proposed fabrication strategy enabled the assembly of individual, diverse emissive microspherical resonators into a unified CROW with ordered architectures of varying color sequences, achieved through precise AFM cantilever tip-based micromanipulation techniques. This structural union of microspherical resonators of varying lengths enabled the manipulation of the entire assembly as a singular unit, offering greater operational precision and ease in micromanipulation operations. The proposed strategy, as revealed by our findings, can be applied to design adaptable CROWs for complex photonic circuits, enabling novel light guiding and color modulation while providing tunable optical properties and precise control over structural integrity. The implications of this research are vast, opening new avenues for scalable, low-cost photonic devices that merge flexibility with high performance, positioning polymer microsphere waveguides at the cutting edge of integrated photonics.

## 4. Experimental Section/Methods

*Optimized self-assembly of fluorophore-doped PS microspheres:*

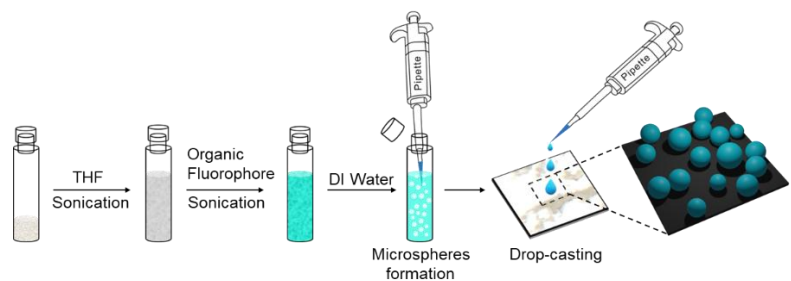

In a standard procedure, 10 mg of PS was dissolved in 4.0 mL of HPLC-grade THF and sonicated for 5 minutes to ensure the complete dissolution of the PS beads. Next, 1 mg of an organic fluorophore was added and thoroughly mixed by sonication for 5 minutes. Subsequently, 1 mL of deionized water was quickly introduced to the mixture, which was left undisturbed for about 10 minutes to allow the formation of microspheres. Finally, 50 µL of the solution was drop-cast onto a clean glass coverslip, and the solvent was allowed to evaporate at room temperature, resulting in fluorophore-doped PS microspheres.

**Supporting Information**

Supporting information is available from the Wiley online library or the authors.

**Acknowledgments**

KY and JH contributed equally to this work. RC acknowledges SERB-New Delhi [SERB-STR/2022/00011 and CRG/2023/003911] for financial support.

TOC

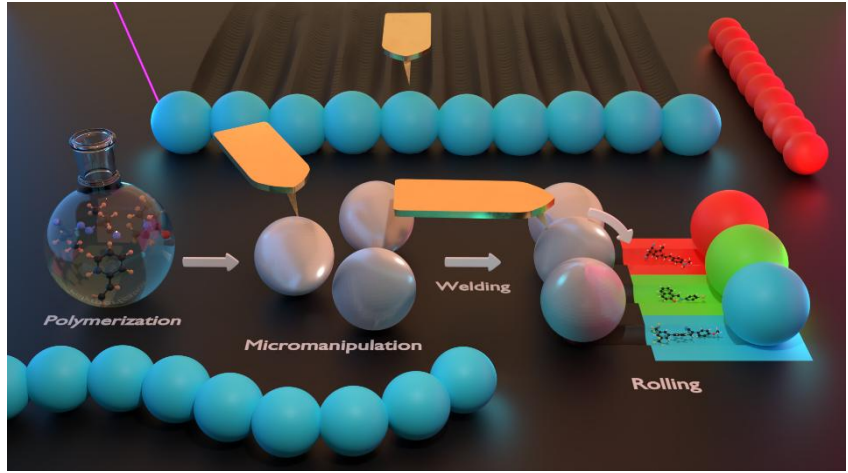